\documentclass{elsart}
\usepackage{psfig} 

\def\apj{ApJ}                 
\def\apjl{ApJ}                
\def\aap{A\&A}                

\usepackage{natbib}
\begin{document}
\runauthor{R\"ottgering et al. } 
\begin{frontmatter}
\title{LOFAR, a new low frequency radio telescope} 
\author[huub]{Huub R\"ottgering}

\address[huub]{Sterrewacht Leiden, PO Box 9513, 2300 RA Leiden, The Netherlands}

\begin{abstract}
LOFAR, the Low Frequency Array, is a large radio telescope consisting
of approximately 100 soccer-field sized antenna stations spread over a region of
400 km in diameter. It will operate at frequencies from $\sim 10$ to
240 MHz, with a resolution at 240 MHz of better than an arcsecond. Its
superb sensitivity will allow for studies of a broad range of astrophysical
topics, including reionisation, transient radio sources and cosmic
rays, distant galaxies and AGNs.  In this contribution 
a status rapport of the LOFAR project and an overview 
of the science case is presented. 
\end{abstract}
\end{frontmatter}

\section{LOFAR, the telescope}

The mission of LOFAR is to survey the Universe at frequencies in the
range 10 - 240 MHz (corresponding to wavelengths of $1.5-30$ m).
Until now this portion of the spectrum has been virtually 
neglected. This is mainly because at these low radio frequencies the
natural image quality is set by ``radio seeing'', i.e. severe image
blurring due to the ionosphere. 
Recent work with the VLA at 74 MHz 
nicely illustrates the problem of ``radio seeing'',
with the apparent position of objects being shifted by more than an
arcminute on timescales of tens of minutes (Cohen, R\"ottgering,
Kassim et al. 2003). \nocite{coh03}  Recently Cotton and Condon (2002)
\nocite{con02} developed an algorithm to correct for these diffraction
effects. The basic idea is that on short regular timescales the
wandering of positions of a number of bright sources within the primary
beam is measured. This timescale is chosen as a compromise between being long
enough so that a sufficient number of sources are visible, and short
enough so that changes in the ionosphere can be smoothly tracked.  For
observations at 74 MHz, typically for 5-10 sources, shifts are measured
on timescales of 1-2 minutes. A Zernike polynomial is then fitted to
the source shifts and the resulting fit is used to correct for the
ionospheric phase fluctuation.

LOFAR's design is such that a similar scheme will work even under the
severe ionospheric conditions present at these low frequencies. Each
of the LOFAR stations will form a baseline with the very sensitive
central core, which comprises 25 \% of the collecting area.  The
sensitivity of each of those baselines is such that, within the
coherence time of the ionosphere and within the primary beam of each
station, enough sources are visible to perform the needed calibration.

\begin{figure}[t]
\centerline{\psfig{figure=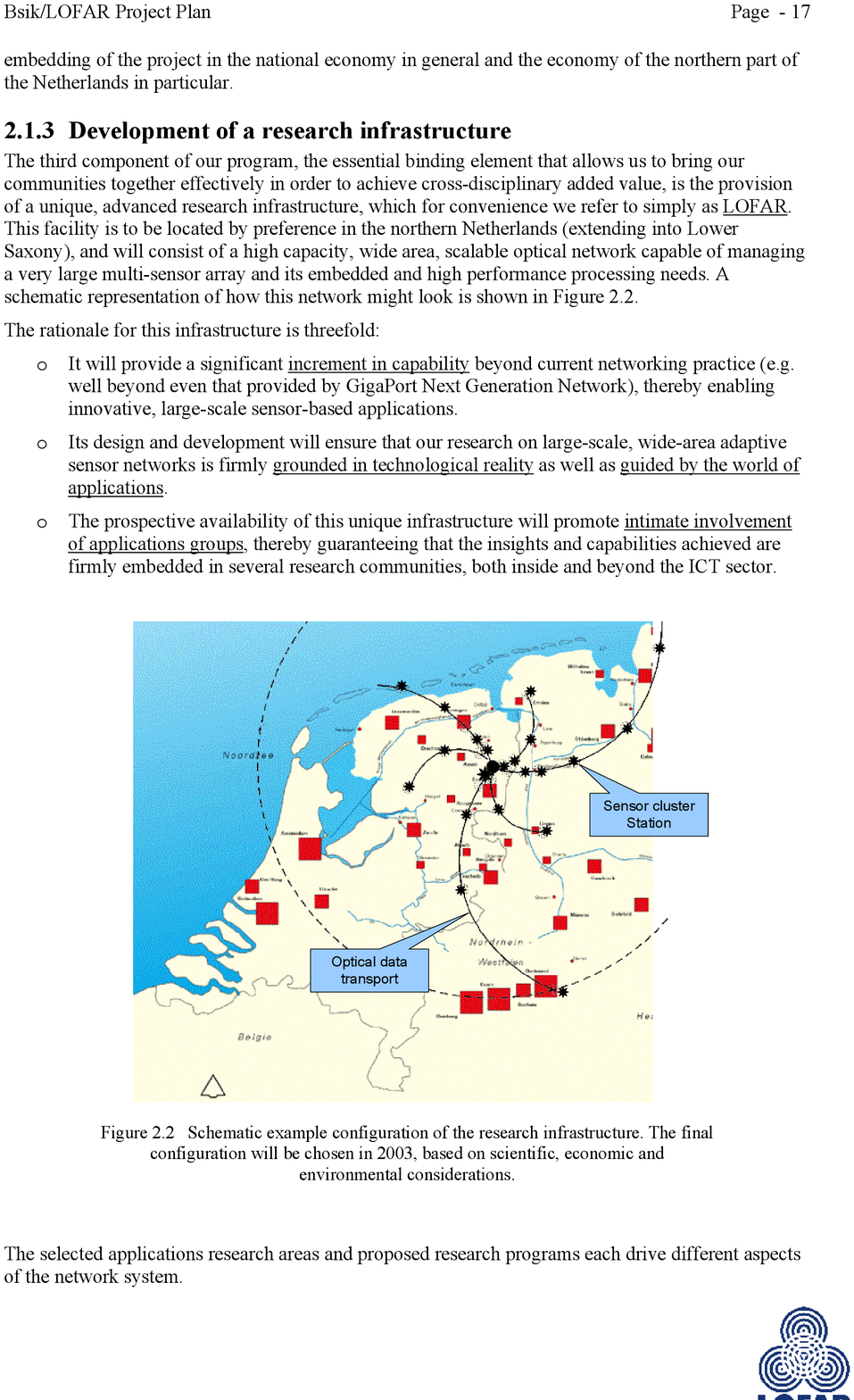,width=0.9\textwidth,clip=}}
\caption{\label{layout} The approximate layout of the LOFAR
stations in a 6-armed log-spiral pattern. Note that this is a
Dutch version of the layout, with similar layouts for the 2 other
proposed sites (see text).}
\end{figure}

LOFAR will have 2 antenna systems, one for the 10-90 MHz range, and
one for the 110-240 MHz range.  These antennas will be placed
in soccer-field sized stations yielding, for each station, effective
aperture sizes  that will range from 50\, m to 150\, m, depending on
frequency.  The signals from each antenna are digitised and fed into
the station beamformer. The beamformer can produce up to eight
coherent ``station-beams'' within the primary power pattern of the
antenna element in use.  In total of the order of 100 stations will
make up the array.  These stations will be distributed over an area
with a diameter of about 400\, km.  
For a sketch of a possible layout if LOFAR is built in the Netherlands,
see Figure 1.

An overview of the resulting resolutions and sensitivities is given in
Table 1. There is an interesting trade-off between sensitivity and
number of beams. If observations are carried out with 8 different
beams, then the maximum bandwidth available for each beam is 4 MHz. It
is also possible to observe with only one beam and use the full 32 MHz 
bandwidth, thus obtaining greater instantaneous sensitivity.

\begin{table}[h]
\caption{An overview of the resolutions and sensitivities of LOFAR for 
a number of fiducial frequencies. The sensitivity are for a single polarisation.} 
\begin{center} 
 \begin{tabular}{|c|c|c|c|}
        \hline \hline
Frequency & Wavelength & resolution   &  Sensitivity \\
 
MHz       &     meter  & arcsec  &  mJy $\left(\frac{t_{\rm int}}{\rm Hour}\right)^{-1/2}
                                         \left(\frac{\rm bandwidth}{4 {\rm MHz}}\right)^{-1/2}$ \\
\hline 
10        &      30    & 15      &  3     \\
30        &      10    &  5      &  1.6   \\
75        &      4     &  2      &  1.0   \\
120       &      2.5   &  1.3    &  0.13  \\
200       &      1.5   &  0.8    &  0.03  \\
\hline
\hline
\end{tabular} 
\end{center} 
\end{table}

\section{Science case} 
The science case for LOFAR is broad; there are important application
of LOFAR not only for astrophysics, but also for studies of the
Earth's ionosphere and the physical properties of the solar wind.
Here we will briefly sketch the general astrophysical 
science case and subsequently
those applications that are most relevant for this workshop.

As a direct result of the calibration scheme that measures phase
fluctuation of the ionosphere, LOFAR will map its dynamic structure
and variability over a wide range of scale sizes.  Emission of a
longwavelength radar directed at the sun would be scattered back by
Solar Coronal mass ejections (CMEs). The Doppler shifts introduced by
different parts of an outward moving CME will result in a
characteristic frequency and time dependent signature.  This not only
enables studies of the CMEs, but would also lead to accurate
predictions for occurrences of geomagnetic storms.

An interesting application of LOFAR is its use as a detector for 
high-energy cosmic rays (HECRs). The existence of HECRs at energies
between 10$^{15}$ -- 10$^{20.5}$ eV is an outstanding challenge for
particle astrophysics. Both the sites and processes for accelerating
these particles are unknown.  A primary CR induces a particle cascade in the
atmosphere which emits coherent radio emission in the terrestrial
magnetosphere (Falcke and Gorham 2003). \nocite{fal03} From the
arrival times and intensities of the radio pulse at various antennas
of LOFAR, the poorly understood development of the electromagnetic
part of the cascade can be studied.  Furthermore, the direction of the
primary particle can then be determined to an accuracy of 1 degree,
potentially revealing the origin of the cosmic rays.

LOFARs large instantaneous beam allows for the first time a sensitive
unbiased survey for radio transients on a variety of time scales,
ranging from a few tenths of seconds to many days. Rapid follow up
with LOFAR at high resolution will give the accurate positions required for
optical and X-ray identifications.  There are numerous classes of
sources that are variable or expected to be variable at low
frequencies and these include radio supernovae, Gamma-ray burst
afterglows, Galactic black-hole/neutron-stars, exo-planets and radio
flare stars.

One of the most exciting goals of LOFAR will be to chart the end of
the Dark Ages when the first stars and AGNs started to ionise the
neutral baryonic gas pervading the Universe.  LOFAR will study at which 
redshift range the bulk of the HI became ionised. Through studies of
the spatial distribution of both the heated and still cold IGM, it
will be possible to 
determine which objects or processes are responsible for
re-ionising the Universe.

\subsection*{Radioloud AGN} 

One of the main goals of LOFAR will be to survey the whole accessible
sky at a number of the lower frequencies (e.g. 15, 30, 75 MHz) down to
the confusion limit.  These surveys will be complimented by surveys at
higher frequencies (e.g. 120 and 200 MHz) with the aim of (i)
obtaining good positional information, essential for optical
identification and 
(ii) determining the higher frequency part of the radio spectrum. 
 
Due to the low observing frequencies, LOFAR surveys will yield large
numbers of radio galaxies with very steep radio spectra. Using the
empirical correlation between radio spectral steepness and distance,
(e.g. de Breuck et al. 2000), \nocite{bre00a} 
LOFAR surveys will be used to efficiently pick out 
very distant ($z>5$) radio galaxies.
As discussed in many contributions to this workshop, 
study of these distant radio galaxies at other
wavelengths will provide information about the formation of massive
galaxies, AGN and proto-clusters. 

It is possible that some of these radio galaxies are located at an
epoch before reionisation has completely occurred. This would open up
the possibility of studying the epoch of reionisation through
observations of the absorbing neutral gas against these very distant
radio galaxies (Carilli et al. 2002) \nocite{car02b}
\subsection*{Starforming galaxies} 
With its unprecedented sensitivity to non-thermal radio emission from
star formation, LOFAR will detect large numbers of {\it star-forming
galaxies} at an epoch at which the bulk of galaxy formation is
believed to occur.  Observing at 200 MHz, LOFAR should be able to
detect the nearby star-forming galaxies, such as M82  and
the ``ultra-luminous infrared galaxy'' Arp 220 (z=0.018), out to
redshifts of respectively $z = 1.1$ and out to $z=3.3$ (e.g. Garrett
2002). \nocite{gar02} Since the ratio of radio flux to sub-mm flux is
a sensitive redshift indicator (Carilli and Yun 1999), \nocite{car99}
LOFAR surveys, in combination with data from new far-IR and millimeter
facilities such as SIRTF, ALMA, and JWST, will therefore provide
distances and thus allow for a complete census of the cosmic
star-formation history, unhindered by the effects of dust obscuration.
\subsection*{Cluster radio halos} 
Due to their large extent, low surface brightness, and
steep-spectra, diffuse cluster sources are difficult to detect with
conventional facilities, such as Westerbork and the VLA.
However, their properties are well
matched to LOFAR's observational capabilities.
The number of cluster halos that could potentially be detected with
LOFAR has been estimated using a simple model (En{\ss}lin and
R\"ottgering  2002). This model takes into account the locally observed
fraction of cluster radio halos, the observed relation between radio and X-
ray luminosity, and a Press-Schecter description of the merging rate of
massive clusters as a function of redshift. A LOFAR survey at 120 MHz 
covering half
the sky to a 5-sigma flux limit of 0.1 mJy (1 hour per pointing) is
feasible on the timescale of one year and could detect of the order of 
1000 halos at the 10
sigma level, of which 25 \% are expected to be at redshifts larger than $z \sim
0.3$.

The XMM X-ray telescope, the Planck satellite and the Sloan Digital
Sky Survey should catalogue as many as 500,000 new clusters
(Barthelmann and White 2002). \nocite{bar02} 
LOFAR has sufficient sensitivity to
detect existing radio halos in all such clusters.
This will 
be very relevant for (i) understanding the dynamics of the cluster gas, 
(ii) determining the origin of their magnetic field content, and (iii)
constraining physical models for the origin of these sources.

\section{The Project}
ASTRON (Dwingeloo, the Netherlands), M.I.T. (Cambridge, USA) and Naval
Research Lab (Washington, USA) are responsible for the design,
construction, operation and software of the LOFAR telescopes.  The
agreed timescale is ambitious and has the aim of full operations in
2008. Important milestones towards this goal are (i) a first test
station with 100 antennas in 2003, (ii) the LOFAR core operational in
2005, and (iii) operations of the central core plus first outer stations in
2006.

A site characterisation committee is presently obtaining data
needed to assess the suitability of LOFAR siting in the
Netherlands, south-west Australia and the southern USA (Texas and New
Mexico).  A decision on the location for LOFAR is expected  to be made 
in 2003. 

The international project is supervised by an International Steering
Committee (ISC), consisting of directors of the participating
institutes. An Engineering Consortium (EC) is responsible for the
design and implementation of the instrument.  The Science Consortium
Board (SCB) is responsible for developing the science case for LOFAR
in close collaboration with the community and gives scientific input
to the EC.

The Science Consortium Board very much welcomes suggestions for
improving or optimising the design of the instrument for general or
very specific applications.  For further information, visit: www.LOFAR.org

{\it Acknowledgements} 
I would like to thank the following of people
for many interesting discussions and fruitful collaborations: Ger de
Bruyn, Aaron Cohen, Bill Cotton, Rob Fender, Michiel van Haarlem,
Melanie Johnston-Hollitt, Namir Kassim, Jan Kuijpers, George Miley,
Jan Noordam and Marco de Vos.


\begin{thebibliography}{}

\bibitem{bar02}
{Bartelmann}, M. and {White}, S.~D.~M.: 2002,
\newblock {\em \aap} {\bf 388}, 732

\bibitem{car02b}
{Carilli}, C.~L., {Gnedin}, N.~Y., and {Owen}, F.: 2002,
\newblock {\em \apj} {\bf 577}, 22

\bibitem{car99}
{Carilli}, C.~L. and {Yun}, M.~S.: 1999,
\newblock {\em \apjl} {\bf 513}, L13

\bibitem{coh03}
{Cohen}, A.~S., {R{\" o}ttgering}, H.~J.~A., {Kassim}, N.~E., {Cotton}, W.~D.,
  {Perley}, R.~A., {Wilman}, R., {Best}, P., {Pierre}, M., {Birkinshaw}, M.,
  {Bremer}, M., and {Zanichelli}, A.: 2003,
\newblock {\em \apj} {\bf 591}, 640

\bibitem{con02}
Condon, W. and Condon, J.~J.: 2002,
\newblock URSI General Assemby, 17-24 August, 2002, Maastricht, The
  Netherlands, paper 0944, pp1-4

\bibitem{bre00a}
{De Breuck}, C., {R{\"o}ttgering}, H., {Miley}, G., {van Breugel}, W., and
  {Best}, P.: 2000,
\newblock {\em \aap} {\bf 362}, 519

\bibitem{fal03}
{Falcke}, H. and {Gorham}, P.: 2003,
\newblock {\em Astroparticle Physics} {\bf 19}, 477

\bibitem{gar02}
Garrett, M.~A.: 2002,
\newblock in {\em Proceedings of the 6th European VLBI Network Symposium, Ros,
  E., Porcas, R.W., Lobanov, A.P., \& Zensus, J.A. (eds.), MPIfR, Bonn,
  Germany, astro-ph/0206270}

\end{thebibliography}

\end{document}